\documentclass[reprint,twocolumn,superscriptaddress,amsmath,longbibliography,amssymb,prb]{revtex4}
\usepackage{graphicx}% Include figure files
\usepackage{dcolumn}% Align table columns on decimal point
\usepackage{bm}% bold math
\usepackage{multirow}
\usepackage{array}
\usepackage{microtype}
\usepackage{soul}

\usepackage[dvipsnames]{xcolor}
\definecolor{nblue}{rgb}{0.0, 0.0, 1.0}
\definecolor{magenta}{rgb}{0.79, 0.08, 0.48}
\usepackage[colorlinks,linkcolor=nblue,urlcolor=magenta,citecolor=magenta,plainpages=false,pdfpagelabels,breaklinks]{hyperref}

\newcommand{\beq}{\begin{equation}}
\newcommand{\eeq}{\end{equation}}
\newcommand{\bea}{\begin{eqnarray}}
\newcommand{\eea}{\end{eqnarray}}

\begin{document}
\title{Abnormal Normal State and Pressure-driven Reentrant Superconductivity in the Heavy $d$-electron Superconductor Rh$_{17}$S$_{15}$}

\author{Xiaofeng Xu}
\email[]{xuxiaofeng@zjut.edu.cn}
%\thanks{These authors contribute equally to this work.}
\affiliation{School of Physics, Zhejiang University of Technology, Hangzhou 310023, China}

\author{J. Y. Nie}
%\thanks{These authors contribute equally to this work.}
\affiliation{State Key Laboratory of Surface Physics, Department of Physics, Fudan University, Shanghai 200438, China}

\author{C. Q. Xu}
%\thanks{These authors contribute equally to this work.}
\affiliation{School of Physical Science and Technology, Ningbo University, Ningbo 315211, China}

\author{Z. M. Zhu}
%\email[]{zimingzhu@hunnu.edu.cn}
\affiliation{Key Laboratory of Low-Dimensional Quantum Structures and Quantum Control of Ministry of Education, Department of Physics
and Synergetic Innovation Center for Quantum Effects and Applications, Hunan Normal University, Changsha 410081, China}

\author{Xiangzhuo Xing}
\email[]{xzxing@qfnu.edu.cn}
\affiliation{Laboratory of High Pressure Physics and Material Science (HPPMS), School of Physics and Physical Engineering, Qufu Normal University, Qufu 273165, China}

\author{Y. L. Huang}
\affiliation{School of Physics, Zhejiang University of Technology, Hangzhou 310023, China}

\author{C. T. Zhang}
\affiliation{Laboratory of High Pressure Physics and Material Science (HPPMS), School of Physics and Physical Engineering, Qufu Normal University, Qufu 273165, China}

\author{N. Zuo}
\affiliation{Laboratory of High Pressure Physics and Material Science (HPPMS), School of Physics and Physical Engineering, Qufu Normal University, Qufu 273165, China}

\author{C. C. Zhao}
\affiliation{State Key Laboratory of Surface Physics, Department of Physics, Fudan University, Shanghai 200438, China}

\author{Z. Y. Zhang}
\affiliation{School of Physics, Zhejiang University of Technology, Hangzhou 310023, China}

\author{W. Zhou}
\affiliation{School of Electronic and Information Engineering, Changshu Institute of Technology, Changshu 215500, China}

\author{W. H. Jiao}
\affiliation{School of Physics, Zhejiang University of Technology, Hangzhou 310023, China}

\author{S. Xu}
\affiliation{School of Physics, Zhejiang University of Technology, Hangzhou 310023, China}
\affiliation{School of Physics, Zhejiang University, Hangzhou 310058, China}

\author{Q. Zhang}
\affiliation{Key Laboratory of Low-Dimensional Quantum Structures and Quantum Control of Ministry of Education, Department of Physics
and Synergetic Innovation Center for Quantum Effects and Applications, Hunan Normal University, Changsha 410081, China}

\author{Zhu-An Xu}
\affiliation{School of Physics, Zhejiang University, Hangzhou 310058, China}

\author{X. B. Liu}
\email[]{xiaobing.phy@qfnu.edu.cn}
\affiliation{Laboratory of High Pressure Physics and Material Science (HPPMS), School of Physics and Physical Engineering, Qufu Normal University, Qufu 273165, China}

\author{Dong Qian}
\affiliation{Key Laboratory of Artificial Structures and Quantum Control (Ministry of Education),
Shenyang National Laboratory for Materials Science, School of Physics and Astronomy,
Shanghai Jiao Tong University, Shanghai 200240, China}
\affiliation{Tsung-Dao Lee Institute, Shanghai Jiao Tong University, Shanghai 200240, China}

\author{Shiyan Li}
\email[]{shiyan$_$li@fudan.edu.cn}
\affiliation{State Key Laboratory of Surface Physics, Department of Physics, Fudan University, Shanghai 200438, China}

\date{\today}

\begin{abstract}
Superconductivity beyond the conventional Bardeen-Cooper-Schrieffer (BCS) framework often emerges out of a normal state that is accompanied by exotic magnetism and thereby displays many exceptional transport and thermodynamic properties. Here we report that the normal state of the heavy $d$-electron superconductor Rh$_{17}$S$_{15}$ is characterized by a weak \textit{ferromagnetism} that persists up to room temperature. We show that the broad hump in its resistivity likely results from the Kondo interaction of the conduction electrons with this novel magnetism. By applying pressure, superconductivity is fully suppressed first. In the high-pressure regime, however, we observe a second dome of superconductivity with its maximum $T_c$ greater than the ambient pressure value, highlighting the possible \textit{unconventional} superconductivity in this heavy $d$-electron sulfide.
\end{abstract}

\maketitle

Understanding the normal state properties of a superconductor, sometimes more obscure than the superconducting state itself, is a key step towards the correct microscopic theory for its superconductivity and phase diagram~\cite{Hussey-Review,Varma-RMP,Hussey-Science-LSCO,Hussey-Science-Rev,Boebinger-Science}. For example, the anomalous normal state of high-temperature copper-oxide superconductors near the optimal doping, characterized by a linear-in-$T$ resistivity all the way up to the melting point, with a slope defined by the Planckian time, has confounded physicists for decades~\cite{Zaanen-review}. In heavy-fermion superconductors, the Kondo lattices comprised of localized $f$ moments are antiferromagnetically screened by the conduction electrons via the Kondo interaction (see the schematic in Fig. 1(a)). Below the Kondo coherence temperature, a Landau Fermi liquid with a flattened band and hence vastly enhanced carrier effective masses emerges~\cite{LuXin-review}, from which Cooper pairs consisting of these heavy electrons form~\cite{steglich-PRL}.

\begin{figure}
\begin{center}
\includegraphics[width=\columnwidth]{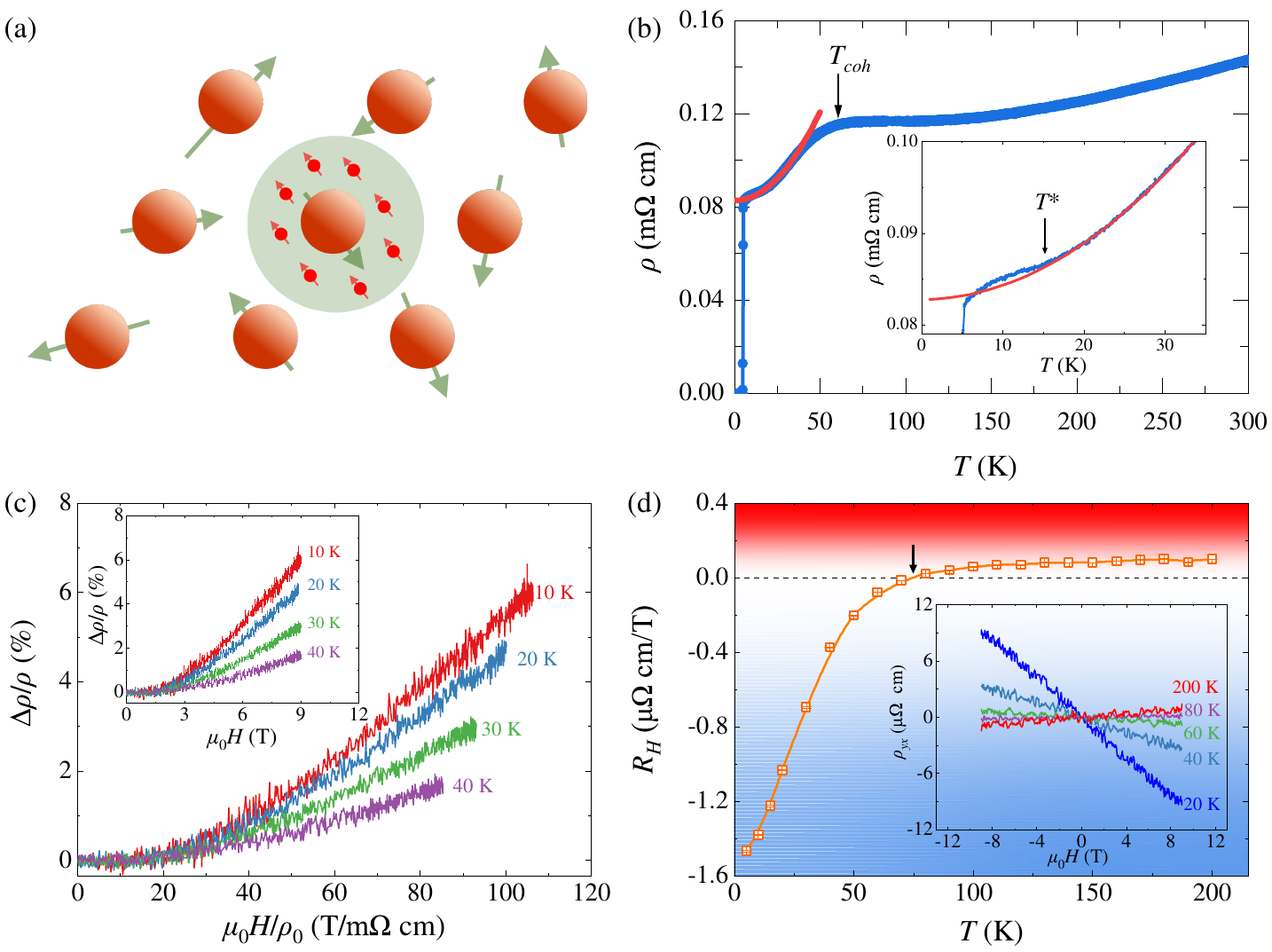}
\caption{(a) A schematic of the Kondo interaction. Here the local moments originate from the $d$-electrons of Rh ions that are screened by the conduction electrons. (b) Temperature dependence of the resistivity in zero-field, which displays a broad hump around $T_{coh}$$\sim$60 K, reminiscent of Kondo coherence, as well as a sharp superconducting transition at 5.2 K. The inset enlarges $\rho$($T$) in the low-temperature regime and the red line shows $\rho$($T$)=$\rho_0$+$A$$T^2$ below 40 K and the deviation below $T^*$$\sim$15 K. (c) The MR (inset) and the Kohler's plot at some representative temperatures, with the current flowing in the plane and the magnetic field orientated along the [111] direction (perpendicular to the plane). (d) The Hall coefficient $R_H$=$\frac{\rho_{yx}}{B}$ extracted from the Hall resistivity $\rho_{yx}$ (inset), which changes sign around 70 K. In the Hall measurements, the current and the field are in the same configurations as in the MR measurement.} \label{Fig1}
\end{center}
\end{figure}

A generic feature of unconventional superconductivity, including the high-$T_c$ cuprates, iron-based superconductors, organic salts as well as heavy-fermion superconductors, is the close proximity to magnetic instabilities in the phase diagram~\cite{Stewart-RMP,Stewart-RMP-HeavyFermions,organic-PRL,Zaanen-review}. Notably, the magnetic order, typically antiferromagnetism in heavy-fermion compounds, can be progressively suppressed by external tuning parameters to a quantum critical point, in the vicinity of which a superconducting dome emerges, leading to the conjecture that the superconductivity is mediated by the quantum fluctuations of this magnetism~\cite{QimiaoSi-NatPhys}. On these grounds, understanding the intricate relationship between superconductivity and magnetism may provide key clues for resolving the enduring mystery of unconventional superconductivity, but the underlying mechanism remains the subject of intense debate.

Superconductors known to exist as natural minerals are extremely scarce, with the rhodium sulfide Rh$_{17}$S$_{15}$ being one of a few examples~\cite{RhS-PRL,RhS-Prozonov,RhS-Daou-JPCM, RhS-Fukui-JPSJ,RhS-Naren-SuST,RhS-NMR,RhS-PRL,RhS-PRL,RhS-Prozonov}. The superconducting miassite Rh$_{17}$S$_{15}$, a mineral found in the Miass river from which its name originates, has the highest $T_c$ ($\sim$5-5.4 K) among the naturally existing superconducting minerals. Besides this uniqueness, its physical properties are remarkable in many aspects, both in its superconducting and normal states. The superconducting transition is characterized by a large heat capacity anomaly at $T_c$ ($\Delta C$/$\gamma T_c$=2), significantly larger than 1.43 from weak coupling BCS theory; the upper critical field $H_{c2}$ was determined to be 20.5 T, a factor of 2 larger than the usual Pauli paramagnetic limit~\cite{zuo-PRB,Hussey-PRL2012}; a line node in its superconducting order parameter was inferred from the $T$-linear dependence of the London penetration depth~\cite{RhS-Prozonov}, suggesting possible unconventional superconductivity in this naturally occurring superconductor, although this claim has been challenged by recent thermal conductivity measurements~\cite{RhS-ShiyanLi-arxiv}. Moreover, its vortex state has been revealed to be highly unconventional, presumably due to some \textit{hidden} magnetic order or a competing superconducting channel~\cite{RhS-Prozonov-arxiv}. Its normal state, on the other hand, is even more exotic. The heat capacity in the normal state uncovers a significantly enhanced electronic contribution, $\gamma$=105 mJ/mol K$^2$, a factor of 5 larger than that from band calculations~\cite{RhS-Naren-SuST}. As a comparison, $\gamma$$\sim$20 mJ/mol K$^2$ in the isostructural homologue Pd$_{17}$Se$_{15}$~\cite{RhS-Naren-SuST}. This greatly enhanced $\gamma$ term, of as yet unknown origin, suggests putative strong electronic correlations in this simple binary system. Intriguingly, the resistivity of Rh$_{17}$S$_{15}$ displays a broad hump around $\sim$60 K, the origin of which has been correlated with resistivity saturation and similar observations in other binary superconductors~\cite{RhS-PRL,Allen-PhysicaB,Fisk-SolidStateCommun}. It was also found that the Hall coefficient changes sign on the border of this resistivity anomaly. Finally, reminiscent of many strongly correlated electron systems, the ratio between the Pauli spin susceptibility $\chi_P$ and the Sommerfeld coefficient $\gamma$, known as the Wilson's ratio, was found to be of order of 2~\cite{RhS-PRL}. All these distinctive characteristics substantially deviate from the expectations for weakly correlated metals and conventional BCS superconductors and thereby enshrine Rh$_{17}$S$_{15}$ in the class of \textit{unconventional} superconductors. However, the origin of all these unconventional properties remains largely enigmatic.

In this Letter, we report a novel, unexpected magnetism in the normal state of the superconducting miassite Rh$_{17}$S$_{15}$ that may help to unravel the origin of the puzzling physical properties described above. Specifically, it is found that the magnetic susceptibility follows the Curie-Weiss law in the normal state, indicating the existence of a local magnetic moment of 0.16$\mu_B$ per Rh atom. Moreover, the $M$-$H$ isotherms explicitly reveal a hysteresis that indicates the presence of weak ferromagnetism in this superconductor. The puzzling heavy-fermion behaviors in its thermodynamic and transport properties can thus be understood in the framework of Kondo physics in this $d$-electron system. Most intriguingly, in common with some unconventional superconductors, we uncover a second superconducting dome in its high pressure phase diagram, highlighting the very unusual superconductivity in this heavy $d$-electron system.

The single crystal growth, characterization, as well as the experimental methods are presented in Notes 1 and 2 of the Supplemental Material (SM)~\cite{SM} (see also the references therein~\cite{XiaoZL-PRX,EuB6-PRL,EuO-PRB,calculation1,calculation2,calculation3,CeCu6,CeCoIn5,CeCo2Ga8,EuRh2Al8,CeBi2}). Crystallizing in a cubic structure with the space group $Pm$3$m$, the unit cell of Rh$_{17}$S$_{15}$ is displayed and dissected in Note 3 of the SM~\cite{SM}. Its zero-field resistivity, as presented in Fig.~\ref{Fig1}(b), is typical of those reported in the literature; it is metallic at high temperature, followed by a broad hump around $T_{coh}$$\sim$60 K (Note 11 of Ref.~\cite{SM}) and finally, a sharp superconducting transition sets in below $T_c$$\sim$5.2 K. In the literature, $T_c$ ranging from 5.0 K to 5.4 K has been reported, presumably depending on the stoichiometry of sulfur~\cite{RhS-PRL,RhS-Prozonov,RhS-Daou-JPCM, RhS-Fukui-JPSJ,RhS-Naren-SuST,RhS-NMR,RhS-PRL,RhS-PRL,RhS-Prozonov}. A close examination of the resistivity curve, as shown in the inset, demonstrates the $T^2$ resistivity below $\sim$40 K, however, the resistivity shows a noticeable departure from this $T^2$ dependence below $\sim$15 K. We mark this temperature as $T^{*}$.

As shown in Fig.~\ref{Fig1}(c), we measured the transverse magnetoresistance (MR) at several temperatures up to 40 K. It can be seen that this sizable MR is quadratic in field ($\Delta\rho/\rho$$\propto$$B^2$). According to the Kohler's rule, plots of $\Delta\rho/\rho_0$ as a function of $H/\rho_0$ ($\rho_0$ is the zero-field resistivity) at distinct temperatures will collapse onto a single curve~\cite{Luo-Kohler-rule,XiaoZL-PRX}. Interestingly, this scaling rule, although derived from semiclassical Boltzmann theory, was found to be widely obeyed in a large number of materials beyond simple metals~\cite{Kohler-cuprate-PRL,Kohler-IBSC-PRR,Kohler-Topo-PRL}. Its violations have also been reported in some novel materials and were generally ascribed to some unconventional mechanisms, such as unusual phase transitions or emergent new physics~\cite{Hussey-Science,Ong-PRL}. As demonstrated in Fig.~\ref{Fig1}(c), Kohler's rule is found to be strongly violated below $T_{coh}$ in Rh$_{17}$S$_{15}$ (above $T_{coh}$, the MR becomes negligible). Other modified forms of the Kohler's rule were also inspected in Note 4 of the SM and found to be strongly violated too~\cite{SM}, suggesting an unusual metallic state below $T_{coh}$ across which the Hall coefficient $R_H$ changes sign (Fig.~\ref{Fig1}(d)).

\begin{figure}
\begin{center}
\includegraphics[width=\columnwidth]{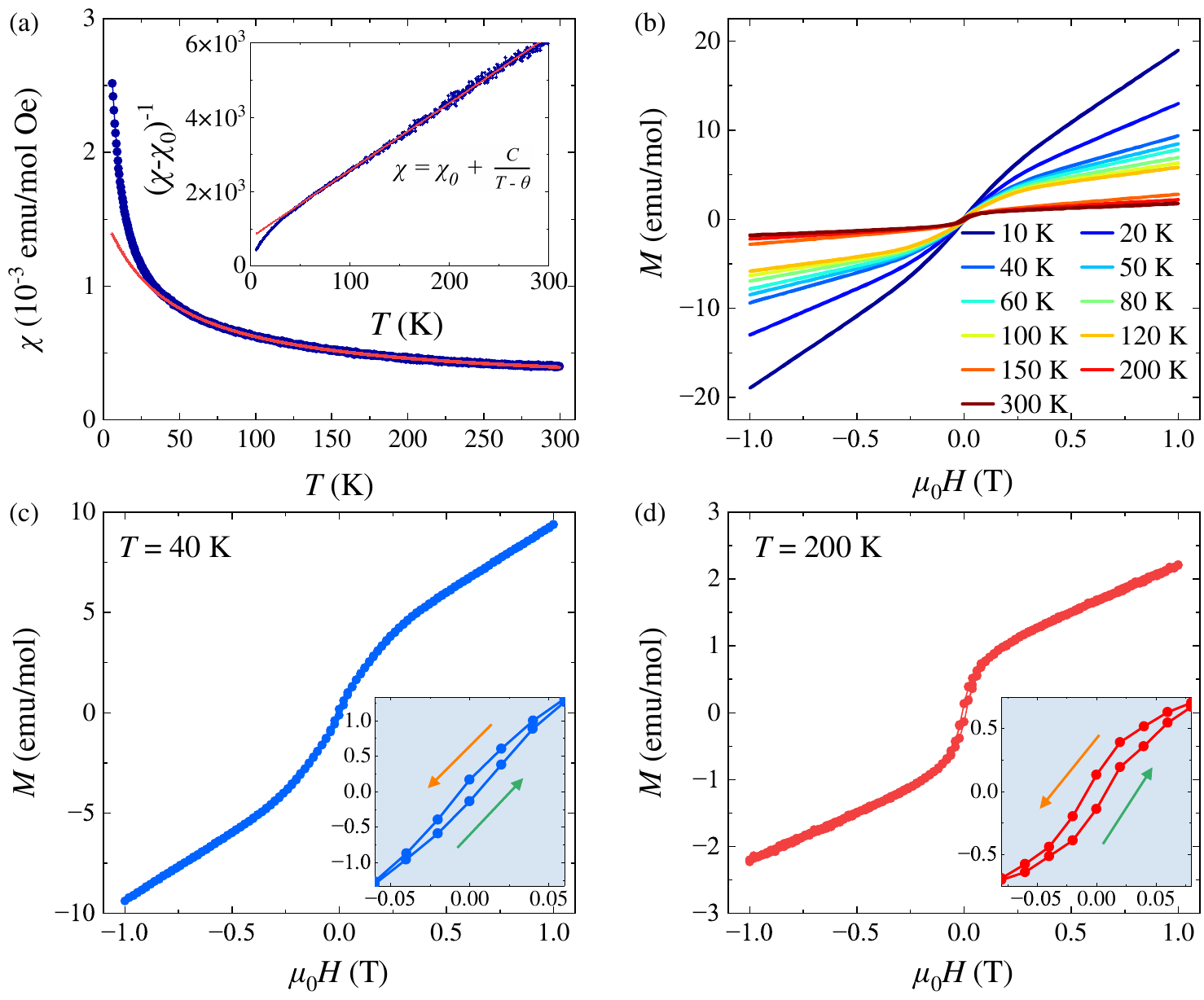}
\caption{(a) The magnetic susceptibility, defined as $\chi$=$\frac{M}{H}$, measured under a magnetic field of 1 T. The red line is the fit to the Curie-Weiss law, $\chi$=$\chi_0$+$\frac{C}{T-\theta}$. The inset further demonstrates this fitting above $\sim$50 K. (b) $M$-$H$ loops at various temperatures below 300 K. (c) and (d) $M$-$H$ loops at representative temperatures 40 K and 200 K, with the low field regime magnified in the insets. The red (green) arrow indicates the field-down (field-up) sweep.} \label{Fig2}
\end{center}
\end{figure}

The magnetic susceptibility $\chi$($T$) was also measured under a 1 T field from room temperature down to $T_c$. As seen in Fig.~\ref{Fig2}, $\chi$($T$) displays a surprisingly strong temperature dependence, similar to that reported in Ref.~\cite{RhS-Naren-JPCM}. We fitted $\chi$($T$) to the Curie-Weiss law, $\chi$=$\chi_0$+$\frac{C}{T-\theta}$, where $C$ is a measure of the effective magnetic moment $\mu_{\textsf{eff}}$, and $\theta$ is the Curie-Weiss temperature. The fitting to this Curie-Weiss law above 100 K yields $\chi_0$ = 2.32$\times$10$^{-4}$ emu mol$^{-1}$ Oe$^{-1}$, $C$ = 0.0553 emu mol$^{-1}$ Oe$^{-1}$ K and $\theta$ = -41.8 K. From the fitted $C$ value and given that the S ions are nonmagnetic, the effective moment per Rh ion can be evaluated to be 0.16$\mu_B$. The negative $\theta$ value -41.8 K indicates the antiferromagnetic interactions between the moments. By extrapolating of the Curie-Weiss law to the lower temperatures, as shown in Fig.~\ref{Fig2}(a), we observe the substantial deviation from this law below $\sim$50 K ($T_{coh}$).

\begin{figure*}
\begin{center}
\includegraphics[width=\textwidth]{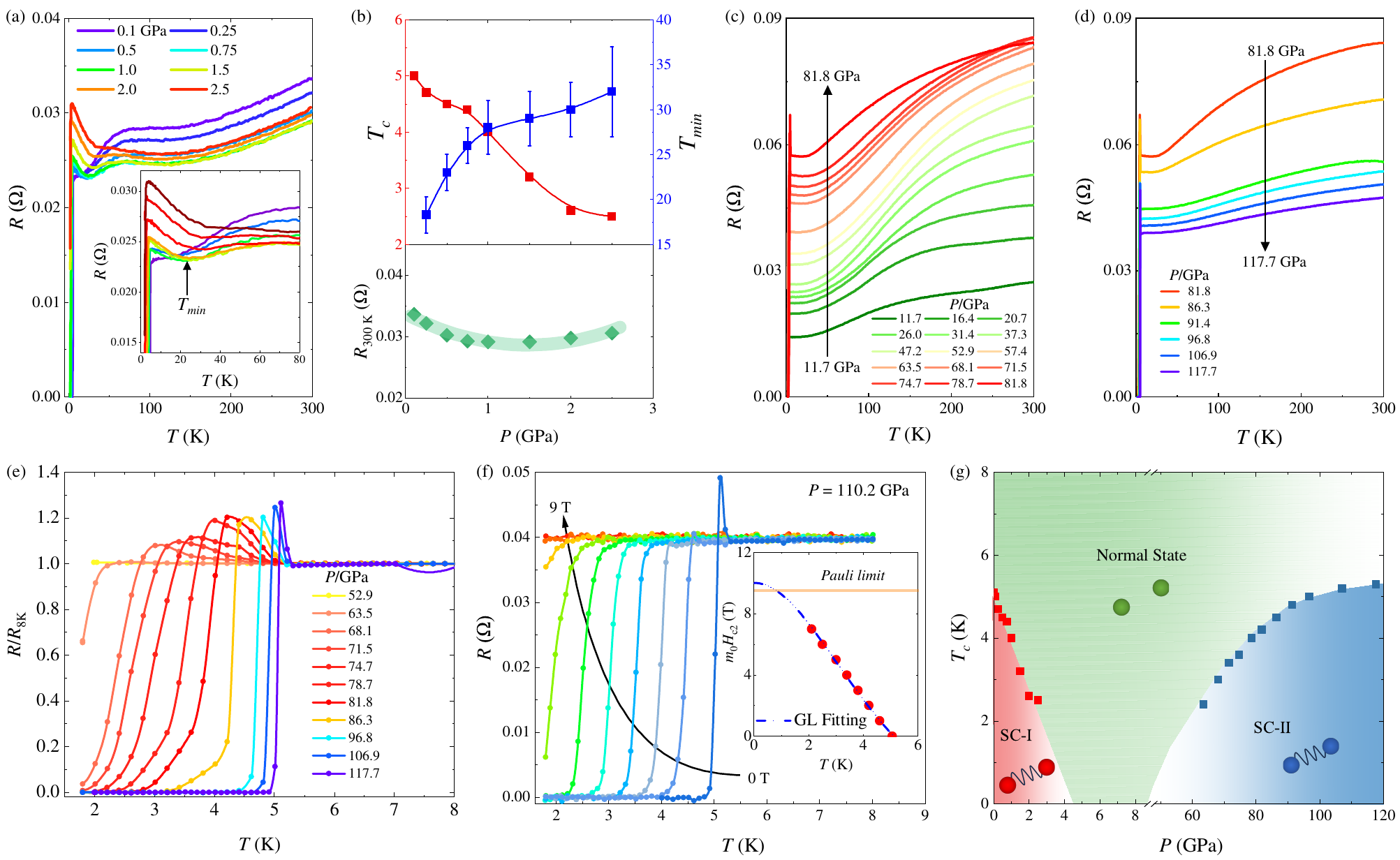}
\caption{(a) Electrical resistance under hydrostatic pressures up to 2.5 GPa, measured in a piston cell. The inset enlarges the low temperature region. (b) Pressure dependence of $T_c$, $T_{min}$ and the room temperature resistance. (c) and (d) The electrical resistance measured under various pressures specified in the legends, performed using a DAC up to 117.7 GPa. (e) The reentrant superconducting transition at high pressures above $\sim$60 GPa. The resistance is renormalized to the corresponding 8 K values for clarity. (f) The upper critical field $\mu_0$$H_{c2}$, determined under $P$=110.2 GPa. The onset temperature is used to determine $T_c$ at each field. The inset shows the temperature evolution of $\mu_0$$H_{c2}$ and the fit to the GL formula, $\mu_0$$H_{c2}$($T$)=$\mu_0$$H_{c2}$(0)$\frac{1-t^2}{1+t^2}$, where $t$=$\frac{T}{T_c}$. (g) The resultant superconducting phase diagram extracted from both piston and the DAC measurements.} \label{Fig3}
\end{center}
\end{figure*}

\begin{figure}
\begin{center}
\includegraphics[width=\columnwidth]{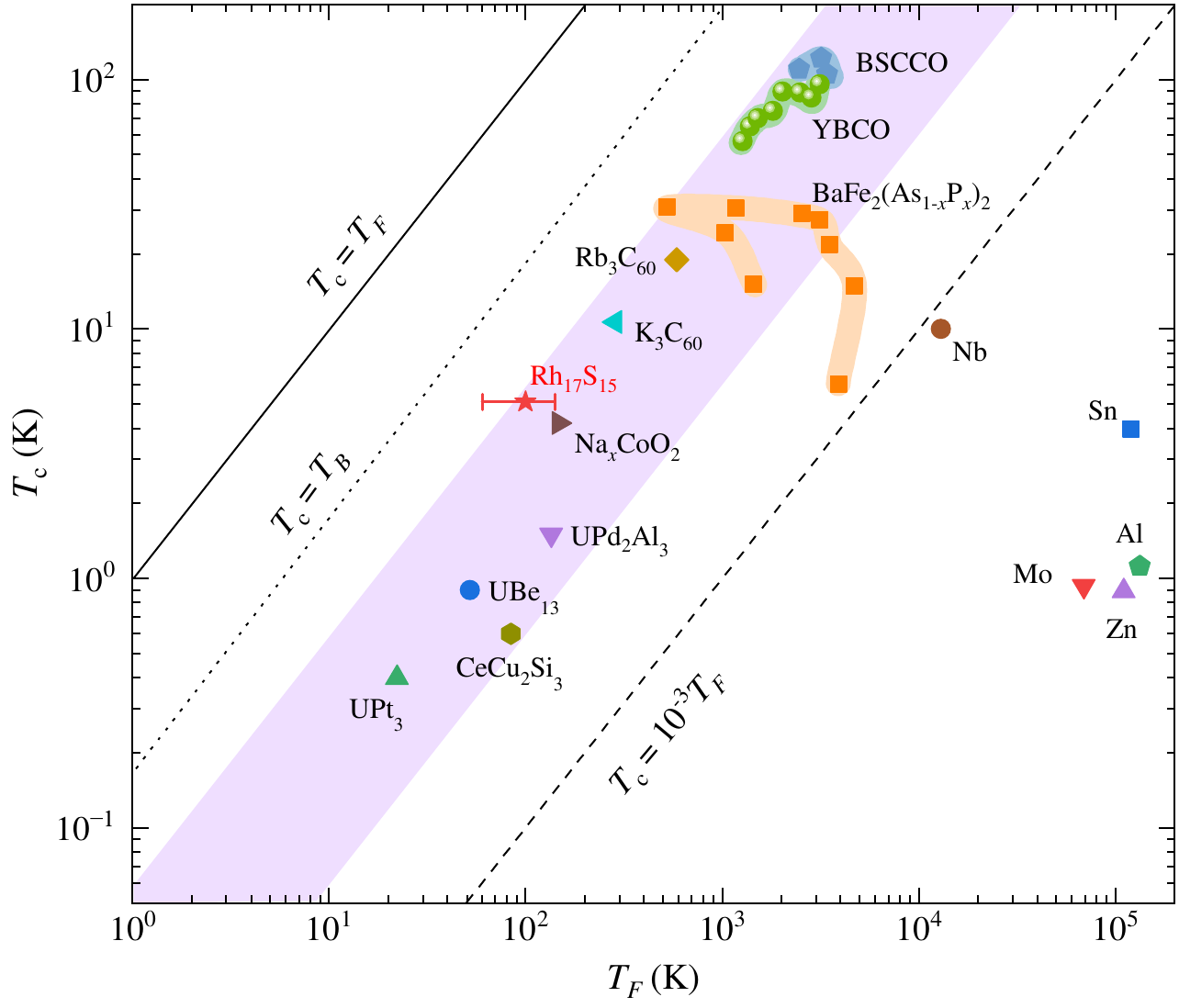}
\caption{The Uemura plot. $T_B$ here is the Bose-Einstein condensation (BEC) temperature for an ideal 3D boson gas, i.e., $T_B$ = 0.0176 $T_F$~\cite{Matsuda-review}.} \label{Fig4}
\end{center}
\end{figure}

The $M$-$H$ profiles present a more striking finding, namely, a ferromagnetic hysteresis in small fields. After subtracting the diamagnetic backgrounds from the sample holder (Note 5 of Ref.~\cite{SM}), as shown in Fig.~\ref{Fig2}(b), the $M$-$H$ loops do not show a linear dependence typically seen in paramagnetic or antiferromagnetic metals, but rather resemble those of ferromagnets. Zooming into the low field regime, we observe clear hysteresis that is sufficiently large to be detected within the sensitivity of the instrument (the hysteresis is 3 orders of magnitude larger than the sensitivity of MPMS used, see Note 5 of Ref.~\cite{SM}). Importantly, a spin glass state can be ruled out as the origin for this hysteresis since no bifurcation in the zero-field-cooled (ZFC) and field-cooled (FC) curves can be resolved (Note 6 of Ref.~\cite{SM}). Conversely, this hysteresis is too large to arise from impurities as we cannot detect any additional elements/phases from XRD and EDX measurements (Note 1 of Ref.~\cite{SM}). See also Ref.~\cite{SM} (Note 1) for sample synthesis procedure in which the high-purity materials of Rhodium granules (99.99$\%$, Alfa Aesar) and Sulfur powder (99.999$\%$, Alfa Aesar) were used. Moreover, perfect diamagnetism is observed below $T_c$, further indicating the high quality of the samples (Fig. S2 of Ref.~\cite{SM}). If the magnetic moments we detected were from other magnetic impurities, the magnetic moment of each impurity atom would be enormously large (assuming one magnetic impurity in 10 formula units, this will lead to an unphysical magnetic moment that is larger than 27$\mu_B$ per impurity). Moreover, we have measured 4 different crystals from different batches and the same results were obtained.

There are many experimental observations that strongly argue against the magnetic impurity as the origin for the magnetic hysteresis and the Curie-Weiss local moments described above. First, in terms of Anderson's theorem, superconducting $T_c$ would be strongly suppressed by magnetic impurities, in both $s$-wave and non-$s$-wave superconductors~\cite{Anderson}. However, the superconducting transition temperature $T_c$ (5.2 K) of our crystals is comparable with those reported in the literature; the onset $T_c$ is 5.4 K in Ref.~\cite{RhS-Prozonov} and 5.0 K in Ref.~\cite{RhS-Daou-JPCM}. Second, the Sommerfeld $\gamma$ term of our samples goes to zero ($T$$\rightarrow$0 K) in the superconducting state and the residual thermal conductivity $\frac{\kappa_0}{T}$$\rightarrow$0 in zero field, indicating no residual quasiparticles and thereby high quality of the samples~\cite{RhS-ShiyanLi-arxiv}. Third, when a small amount of magnetic Fe ($<$ 4$\%$) is doped into the Rh sites, we find that both $T_c$ and the magnetic hysteresis are quickly suppressed (data not shown here). This strongly suggests that the ferromagnetism we observed in Rh$_{17}$S$_{15}$ is not due to magnetic impurities. On these grounds, we conclude that the only possible source of magnetic moments and hysteresis is indeed rooted in the Rh ions. In common with many ferromagnets, the ferromagnetism is observed up to room temperature in Rh$_{17}$S$_{15}$ (Note 7 of Ref.~\cite{SM}).

The anomalous electronic states in Rh$_{17}$S$_{15}$ were further tuned by applying pressure, using both a piston-type pressure cell and a diamond anvil cell (DAC). Figure~\ref{Fig3}(a) shows the resistance curves measured under various pressures up to 2.5 GPa in the piston cell, with the low temperature region enlarged in the inset. With increasing pressure, the broad hump around 60 K is gradually suppressed. Simultaneously, the resistivity upturn below $T_{min}$ becomes more pronounced and the superconducting $T_c$ is gradually suppressed. Similar to $T^*$ at ambient pressure (Fig.~\ref{Fig1}(b) inset), this $T_{min}$ may be associated with the enhanced Kondo coupling between itinerant electrons and the localized moments, or it could mark an additional ordering temperature. The anti-correlation between $T_c$ and $T_{min}$ presented in Fig.~\ref{Fig3}(b) suggests that the energy scales of these two states compete in determining the ground state. The variation of the room temperature resistance with pressure is presented in the lower panel of Fig.~\ref{Fig3}(b); with increasing pressure, it first decreases and then goes up slightly.

The pressure effect on the resistance of Rh$_{17}$S$_{15}$ was further investigated using a DAC that produces a pressure as high as 117.7 GPa. Under $\sim$11 GPa, no superconductivity is observed down to 2 K (Fig.~\ref{Fig3}(c)). With increasing pressure, the resistance is firstly enhanced up to 81.8 GPa (Fig.~\ref{Fig3}(c)), above which it starts to decrease (Fig.~\ref{Fig3}(d)). Surprisingly, superconductivity re-emerges above $\sim$60 GPa and continues to increase with increasing pressure (Fig.~\ref{Fig3}(e)). At 117.7 GPa, the highest pressure measured in this study, $T_c$ reaches 5.3 K, slightly higher than the ambient pressure value. Note that resistance goes up sharply at $T_c$ (Fig.~\ref{Fig3}(e)), typically seen in the high pressure measurements and may be ascribed to the pressure inhomogeneity effect at such high pressures~\cite{Yang-MoP}. It is remarkable to see the complete superconducting transition with absolute zero resistivity at such high pressures, which is unlikely to be ascribed to any impurity phase. The upper critical field $\mu_0$$H_{c2}$ is determined at 110.2 GPa, by sweeping temperature at constant fields, as plotted in Fig.~\ref{Fig3}(f). The Ginzburg-Landau (GL) fitting gives a zero-temperature $\mu_0$$H_{c2}$ value slightly higher than the weak-coupling Pauli paramagnetic limit 1.84$T_c$, as demonstrated in the inset of Fig.~\ref{Fig3}(f). The superconducting phase diagram is summarized in Fig.~\ref{Fig3}(g).

For 3D systems like Rh$_{17}$S$_{15}$, the Fermi temperature $T_F$ is given by $T_F=\frac{\hbar^{2}}{2}(3\pi^2)^{\frac{2}{3}}\frac{n_{s}^{2/3}}{k_Bm^{*}}$, where $n_s$ is the carrier concentration and $m^{*}$ is the effective mass of carriers~\cite{Matsuda-review,Uemura}. As a crude estimate, $n_s$ is evaluated to be 4.16$\times$10$^{20}$/cm$^3$ from the low temperature Hall coefficient. Although the multiband effect as in Rh$_{17}$S$_{15}$ can give rise to an increase of $n_s$, the uncertainty in $n_s$ is not expected to exceed by a factor of 10. The effective mass $m^*$ is estimated to be 30$m_0$ ($m_0$ the free electron mass) from the carrier concentration $n_s$ and the heat capacity $\gamma$ term~\cite{RhS-NMR,RhS-ShiyanLi-arxiv}. On this footing, $T_F$ is estimated to be less than 100 K. As noted from the Uemura plot in Fig.~\ref{Fig4}, Rh$_{17}$S$_{15}$ is well situated in the unconventional regime, suggesting the possible unconventional superconductivity in this system.

We now turn to the origin of the heavy-electron (flat-band) phenomenon and novel physical properties described above, in this structurally simple, $d$-electron system. Unlike the more compact nature of partially filled 4$f$ and 5$f$ electrons, $d$ electrons are more extended and less susceptible to forming localized states and heavy quasiparticles. Heavy-fermion states have been reported in the V pyrochlore-containing oxide, $d$-electron LiV$_2$O$_4$, wherein the magnetic frustration due to its structural motif is broadly believed to play a vital role in their formation~\cite{LiV2O4-PRL,LiV2O4-Coleman-PRL}. Recently, heavy-fermion and strange-metal behaviors have also been reported in the 3$d$ kagome metal Ni$_3$In, where the correlated electron physics can be understood in the context of destructive quantum interference, namely, a hopping interference-driven realization of the Kondon lattice due to its structural network~\cite{Checkelsky-Ni3In}. In this kagome network, the orbitals on the hexagonal plaquette are trapped by the destructive interference of nearby hopping pathways in the triangular plaquette, leading to compact localized states and flat bands~\cite{Checkelsky-Ni3In}. Similar physics was also observed in the pyrochlore metal CaNi$_2$ which contains topological flat bands arising from the 3D destructive interference of electronic hopping~\cite{Checkelsky-CaNi2}. In all these systems, heavy fermions (flat bands) originate from frustration-driven compact localized states in their band structure~\cite{Checkelsky-NatureRevMater}. However, the physics in Rh$_{17}$S$_{15}$ is remarkably different. Rh ions are generally thought to be non-magnetic, such as in LaRhIn$_5$, LaRhSi$_3$, RhSn$_4$ etc., and the structure of Rh$_{17}$S$_{15}$ is cubic and centrosymmetric, and therefore no magnetic frustration is expected. The first-principles calculations provide some clues for understanding this correlated metallic behavior, as shown in the SM~\cite{SM} (Note 9). Indeed, without including the Coulomb interactions of $d$-electrons, no localized moments can be realized. To obtain local moments of the magnitude of 0.1-0.2$\mu_B$, a strong Coulomb interaction of $U$$\sim$ 6 eV must be invoked~\cite{SM}. This Coulomb interaction is rather strong, but lower than those expected in the strongly correlated cuprates and heavy fermion systems. In reality, smaller Coulomb interactions can also lead to a comparable size of local moments, with the aid of atomic defects or inhomogenity~\cite{SM}.

Weak ferromagnetism may occur in antiferromagnets due to the anisotropic exchange interaction, the so-called Dzyaloshinshy-Moriya (DM) effect. This effect is to cant the antiferromagnetic moments by a small angle and then results in a small ferromagnetic component among the overall antiferromagnetically coupled moments. However, this effect is expected to vanish when the crystal field has an inversion symmetry with respect to the center of the two magnetic moments, as in centrosymmetric Rh$_{17}$S$_{15}$. The weak ferromagnetism observed in this work therefore merits further theoretical insights. On the other hand, the reentrant superconductivity has only been observed in a handful of unconventional superconductors where \textit{magnetic fluctuations} play an essential role~\cite{Panagopoulos-review}. The origin for the reentrant superconductivity in Rh$_{17}$S$_{15}$ is unknown; however, it may also be related to the novel magnetism revealed in this study.

In conclusion, we have demonstrated an unusual normal state in the heavy $d$-electron superconductor Rh$_{17}$S$_{15}$. This normal state is characterized by a broad resistivity hump below which the Hall effect changes sign and the (modified) Kohler's rule is strongly violated. We uncovered the novel magnetism in this normal state that manifests a weak ferromagnetic ordering of localized moments, which may help to reframe the understanding of its heavy-fermion behavior and the abnormal physical properties. By applying high pressures, we uncovered a reentrant superconducting phase under high pressure that is typically seen in unconventional superconductors with prominent magnetic fluctuations. These findings therefore raise an exciting prospect to investigate the origin of this novel ferromagnetism and the emergent superconductivity from its intricate many-body interactions.

The authors would like to acknowledge Michael Smidman, Nigel Hussey, Ali Bangura, Xin Lu, Jianhui Dai, Chao Cao, Zengwei Zhu for stimulating discussions. This work was supported by NSFC.

XX., JN., CX. and ZZ. contributed equally to the work.

%\bibliography{RhS}

%References
%\bibliographystyle{apsrev4-1}

\end{document}